\documentstyle[epsf]{ptptex}

\def\Journal#1#2#3#4{{#1}{\bf #2} (#4), #3}
\def\PTP{Prog.~Theor.~Phys.~}
\def\PTPS{Prog.~Theor.~Phys.~Suppl.~}

\def\NPB{Nucl.~Phys. \bf{B}}
\def\PLB{Phys.~Lett. \bf{B}}

\def\PRL{Phys.~Rev.~Lett.~}
\def\PRD{Phys.~Rev. \bf{D}}

\notypesetlogo  
\title{%
Further Analysis on $\sigma$-particle Properties
}

\author{
Shin {\sc Ishida}, 
Taku {\sc Ishida}$^{*}$, Muneyuki {\sc Ishida}$^{**}$,\\
Kunio {\sc Takamatsu}$^{***}$,
and Tsuneaki {\sc Tsuru}$^{*}$
}

\inst{%
Atomic Energy Research Institute, 
College of Science and Technology,\\
Nihon University, Tokyo 101
\
$^{*}$National Laboratory for High Energy Physics (KEK), 
Tsukuba 305
\\
$^{**}$Department of Physics, 
University of Tokyo, Tokyo 113
\\
$^{***}$Department of Engineering, Miyazaki University, 
Miyazaki 988-12
}

\recdate{%
May 25, 1997
}

\abst{%
In a previous work, we have reanalyzed the I=0 S-wave $\pi\pi$-scattering 
phase-shift through a new method of Interfering Breit-Wigner amplitudes,
and have shown the existence of a light scalar, $\sigma$ particle, 
accompanied by a negative ``repulsive core"-type background phase-shift, 
whose origin may have some correspondence to a ``compensating'' $\pi\pi$ 
contact interaction. In this work we make further analysis on the
phase shift under $K\overline{K}$ threshold from the same standpoint, 
to determine precise values of $\sigma$ mass and width, so far as present
experimental data are concerned.
}

\begin{document}

\maketitle
\section{Introduction}\label{sec:int}
Whether a light iso-scalar resonance ($J^{PC}=0^{++}$),
called "$\sigma$"-particle, exists or not is one of the most interesting
and important problems in hadron spectroscopy. Phenomenologically, 
since of its light mass and of its ``vacuum" quantum number, 
it may affect various processes. Theoretically, for example,
in the Nambu-Jona-Lasinio-type models\cite{rf:Sca,rf:HS,rf:HK,rf:KT}, 
realizing the situation of dynamical breaking of chiral symmetry
and believed to be a low-energy effective theory of QCD, 
existence of $\sigma$-meson is predicted 
as a chiral partner of the Nambu-Goldstone $\pi$-boson.
Correspondingly the extensive experimental investigations in the
$\pi\pi$-channel have been made for many years.
The I=0 S-wave $\pi\pi$ phase shift $\delta^0_0$ is now well 
known to rise smoothly to $90^\circ$ at around 900 MeV, then shows a rapid 
step-up by $180^\circ$ near the $K\overline{K}$ threshold, and
reaches only to $270^\circ$ at $m_{\pi\pi}\sim$1200 MeV.
This behavior of $\delta^0_0$ was thought, in the 1976-through-1994 
editions of PDG, to be due mainly to the narrow $f_0(980)$ and the 
broad $f_0(1370)$,
\footnote{
However in the latest edition,\cite{rf:PDG96} the behavior of $\delta^0_0$ is 
understood as due to, in addition to the $f_0(980)$ and $f_0(1370)$,
a very broad $f_0(400\sim 1200)$ or $\sigma$. See also the 
analyses suggesting the $\sigma$ existence,\cite{rf:Ka,rf:Tor}
which introduce a repulsive $\delta_{B.G.}$, similarly as in our case.}
since there remains no phase shift for light $\sigma$-particle.

In contrast with this interpretation,
we have shown in the previous work\cite{rf:pipip} (to be referred as {\bf I}), 
re-analyzing the phase shift $\delta^0_0$ systematically up to 1300 MeV,
a possibility for the existence of ``$\sigma$(555)''-resonance with 
a rather narrow width of a few hundred MeV 
in addition to these two resonances. 

The reasons which led us to a different result from the conventional one, 
even with the use of the same data of phase shifts, are twofold:
On one hand technically, we have applied a new method of
Interfering Breit-Wigner Amplitude (IA method) for the analyses,
where the ${\cal T}$-matrix (instead of ${\cal K}$-matrix in the conventional 
treatment) for multiple resonance case
is directly represented by the respective Breit-Wigner amplitudes
in conformity with unitarity,
thus parametrizing the phase shifts
directly in terms of physical quantities, such as masses and coupling
constants of the relevant resonant particles.

On the other hand physically, we have introduced 
a ``negative background phase'' $\delta_{B.G.}$ of hard core type,
(with a core radius of about pion size,) 
making enough room for $\sigma$-resonance. 
Here it is suggestive to remember that a similar type of $\delta_{B.G.}$ 
is well-known to exist in the $\alpha$ nucleus-$\alpha$ nucleus scattering 
and in the nucleon-nucleon scattering\cite{rf:core}.
A possible origin of $\delta_{BG}$ in the $\pi\pi$ system 
seems\cite{rf:MY} to have some correspondence to the 
``compensating'' repulsive $\lambda\phi^4$ interaction in the NJL-model or 
the linear $\sigma$ model, which is required from the viewpoint 
of the current algebra and the PCAC. 

In this work, we shall extend the re-analysis of the phase shift 
in {\bf I} to determine the mass and coupling constant 
of $\sigma$ and core radius as precise as possible,
from the present experimental data of $\delta^0_0$, 
inspecting especially the data under $K\overline{K}$ threshold.

\section{Applied formulas}\label{sec:app}
For our purpose we shall analyze the $\delta_0^0$ between the 
$\pi\pi$-threshold and an energy slightly below the 
$K\overline{K}$-threshold (980 MeV), taking into account the effects of 
two resonances, $\sigma$ and $f_0(980)$, and of the $\delta_{B.G.}$.
Effects from other higher-mass resonances ($f_0(1370)$, $f_0(1500)$...)
and from other channels than $\pi\pi$ are ignored.

The applied formulas of IA-method (in the case of one channel
with two resonances) are as follows.

The relevant partial S-wave
${\cal S}$ matrix element $S$ in the $2\pi$ system 
is represented by the phase shift $\delta (s)$
and the amplitude $a(s)$.
The $\delta$ is given by the sum of
$\delta^{Res.}$ and $\delta^{B.G.}$, respectively, due to
the resonance and background.
The $\sigma$ and $f_0(980)$ resonances contribute additively to
$\delta^{Res.}$. 
\begin{eqnarray}
S&=& e^{2i\delta (s)}=1+2ia(s) \label{eq:sdef}\\
\delta(s)&=& \delta^{Res.}(s)+\delta^{B.G.}(s),\ \ \ 
\delta^{Res.}(s) = \stackrel{(\sigma)}{\delta}(s)+
\stackrel{(f_0)}{\delta}(s). \label{eq:delb}
\end{eqnarray}
Correspondingly the total ${\cal S}$ matrix is
given by the product of 
individual ${\cal S}$-matrices. 
\begin{eqnarray}
S=S^{Res.}S^{B.G.}=\stackrel{(\sigma)}{S}\stackrel{(f_0)}{S}S^{B.G.}.
\label{eq:tots}
\end{eqnarray}
The unitarity of the total ${\cal S}$ matrix is
now easily seen to be satisfied by the ``unitarity of individual
$S$-matrices".
Each of $S^{Res.}$ is given by corresponding amplitudes
$\stackrel{(R)}{a}$'s, which is taken as the following relativistic 
Breit-Wigner (BW) form
\begin{eqnarray}
\stackrel{(R)}{S}(s) &=& e^{2i \delta^{(R)} }
=1+2i\stackrel{(R)}{a}(s)\ \ :R=\sigma ,\ f_0 \label{eq:bwa}\\
\stackrel{(R)}{a}(s) &=&
\frac{-\sqrt{s}\Gamma_R (s)}
{(s-M_R^2) + i\sqrt{s}\Gamma_R (s)}, \label{eq:bwb}\\
\sqrt{s}\Gamma_R (s)&\equiv& \rho_1 g_R^2,\ \ 
\rho_1 \equiv \frac{p_1}{8\pi\sqrt{s}},\ \ p_1\equiv\sqrt{s/4-m_\pi^2},
\label{eq:bwc}
\end{eqnarray}
where $\Gamma_R (s=M_R^2)$ represents the peak width
$\Gamma_R^{(p)}$ of the resonance $R$,
$g_R$ is the $\pi\pi$-coupling constant,
$\rho_1$ is the $\pi\pi$-state density and $p_1$ is the CM momentum
of the pion.
Here it is to be noted that the total resonance amplitude
$a^{Res.}(\equiv (S^{Res.}-1)/2i)$ is represented by the respective
amplitudes as
\begin{eqnarray}
a^{Res.}=\stackrel{(\sigma)}{a}+\stackrel{(f_0)}{a}
+2i\stackrel{(\sigma)}{a}\stackrel{(f_0)}{a},
\end{eqnarray}
where the last term, looking like an
``interference" of the amplitudes, guarantees our amplitude to
satisfy the unitarity constraint.

The
$\delta_{B.G.}$ is supposed to be of the repulsive hard core type;
\begin{eqnarray}
\delta^{B.G.}(s)=-{\bf p}_1r_c. \label{eq:bga}
\end{eqnarray}
In the actual analysis we have applied, in order to have a global
fit in all relevant energy region, the relativistic BW formula
(\ref{eq:bwb}) with a revised width
\begin{eqnarray}
\tilde{\Gamma}_R(s)&\equiv&\Gamma_R(s)F(s)\hspace{6.5cm} (a),\nonumber\\ 
F(s)&=&1-\frac{s_0}{s+s_0}e^{-s/M_0^2},\ (M_0=400\ {\rm  MeV})\hspace{2cm}(b),
\label{eq:width}
\end{eqnarray}
instead of $\Gamma_R(s)$ defined in Eq.(\ref{eq:bwc}).
As a matter of fact there has been no established form of relativistic
BW formula, and the concrete form of $F(s)$ is considered to be 
determined following the dynamics of strong interactions.
The conventional BW form (\ref{eq:bwb}) ($\tilde{\Gamma}_R(s)$ with 
$F(s)=1$ in Eq.(\ref{eq:width}))gives, in the case of broad width,
an unphysical mass spectrum in the low energy region.
The phenomenologocal form of $F(s)$ with a parameter $s_0$
given in (\ref{eq:width}) is
so designed as
to give a good fit in the low-energy region affecting it
only in the energy region
below 400 MeV. Thus $s_0$ plays a similar
role as a scattering length.

\section{Mass and width of $\sigma$ and core radius}
\label{sec:mas}

\begin{figure}[t]
 \epsfysize=12.5 cm
 \centerline{\epsffile{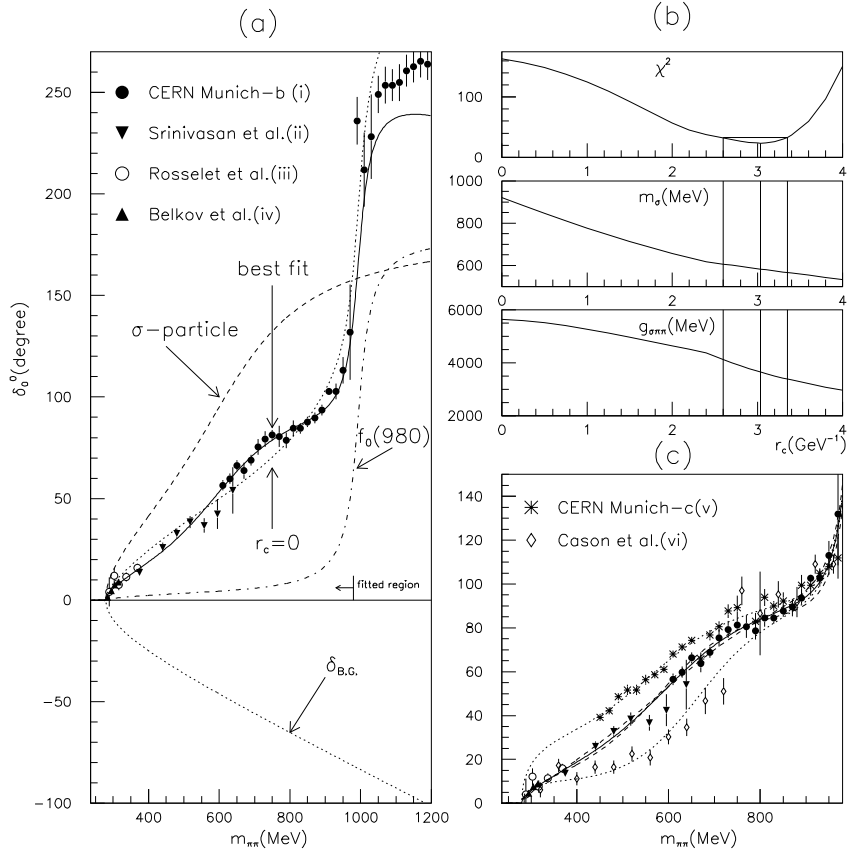}}
\caption{I=0 $\pi\pi$ scattering phase shift.
(a) Best fit to the standard $\delta_0^0$.
The respective contributions to $\delta_0^0$ from $\sigma ,f_0(980)$
and $\delta_{B.G.}$ are also given. The dotted line with label of
``$r_c$=0'' represents conventional fit without 
the repulsive background thus far made.
(b) $\chi^2, M_\sigma$ and $g_\sigma$ versus $r_c$. 
(c) Best fits to the upper and lower $\delta_0^0$ (dashed lines),
and fittings with 3-sigma deviation
to the standard $\delta_0^0$
corresponding to $r_c$=2.6 and 3.35 GeV$^{-1}$ (dotted lines).}
\label{fig:i0}
\end{figure}
The $\delta^0_0$ over $m_{\pi\pi}>600$ MeV is well determined by
the analysis of the CERN-M\"unich experiment.
Among 5 independent analyses performed\cite{rf:I0} in Grayer74, (i)
the one originally presented in Hyams73 (b-analysis in Grayer74)
is widely accepted, partly because of its agreement to those reported
by Protopopescu73\cite{rf:Pr}. 
(ii) Srinivasan75 gave the phase between 350-600 MeV,
(iii) Rosselet77 and (iv) Bel'kov79 determined the phase between 
$\pi\pi$-threshold and 400 MeV.
(i)-(iv) will be used as the ``standard phase shift $\delta_0^0$"
in the present analysis.
However, there are several works to report different behaviors of
$\delta_0^0$ within $\pm 20^\circ$ ambiguities in the 
$m_{\pi\pi}$=400-800 MeV region.
We will utilize them to estimate the allowed region of 
$m_\sigma$ and $g_\sigma$ in our analysis :
(v) The c-analysis of Grayer74 
and (vi) Cason82 are taken as the upper and lower bounds
of $\delta_0^0$, respectively.

The properties of $f_0(980)$ have already been 
obtained in {\bf I} by analyzing systematically the
standard $\delta_0^0$, elasticity, and
$\pi\pi\rightarrow K\overline{K}$ phase shift together
in the range of $m_{\pi\pi}$=600 $\sim$ 1300 MeV,
with the 2-channel ($\pi\pi$, $K\overline{K}$) 
three resonance ($\sigma$, $f_0(980)$, $f_0(1370)$) formula in
the IA method. The obtained values are 
$M_{f_0}$=993.2$\pm$6.5$_{st}$$\pm$6.9$_{sys}$ MeV and 
$g_{f_0}$=1680$\pm$91 MeV, which are used as fixed
parameters in the following analysis.
\footnote{The $f_0(980)$ does not play essential role in
the relevant mass region since of its small coupling (width).}

\begin{table}[t]
\caption{Obtained values of parameters and their errors.
$g^r\equiv g\cdot
F^{1/2}(s=M_R^2)$.
As for the definitions of $\Gamma^{(p)}$ and $\Gamma^{(d)}$,
see the text.}
\begin{center}
\begin{tabular}{l|ccc}
\hline
 & standard $\delta_0^0$ (i-iv) & upper bound (v) & lower bound (vi)\\
\hline
$M_\sigma$     &   585$\pm$20 MeV  & 535 MeV & 650 MeV\\
$g_{\sigma\pi\pi}^r$  & 3600$\pm$350 MeV & 3670 MeV & 3465 MeV\\  
$\Gamma_{\sigma\pi\pi}^{(p)}$ & 385$\pm$70 MeV & 430 MeV & 335 MeV\\
$\Gamma_{\sigma\pi\pi}^{(d)}$ &  340$\pm$45 MeV     & 370 MeV   &  315 MeV \\ 
$r_c$    &   3.03$\pm$0.35 GeV$^{-1}$ & 3.03(fixed) & 3.03(fixed)\\ 
 &0.60$\pm$0.07 fm& (0.6 fm) & (0.6 fm)\\
\hline
\end{tabular}
\label{tab:mw}
\end{center}
\end{table}
Fig.2(a) shows the result of the best fit (solid line)
to standard $\delta_0^0$ by the formula (\ref{eq:delb}) given above,
which includes the sum of three contributions: $\sigma$, $f_0(980)$, and 
$\delta_{B.G.}$.
The best-fitted values of parameters are 
$M_\sigma$=585 MeV, $g_\sigma$=3.6 GeV
(corrsponding to decay width of 340 MeV), and 
$r_c$=3.03 GeV$^{-1}$(0.60 fm)\footnote{ 
The significant difference of the value of 
$M_\sigma$ from that($553.3\pm 0.5$ MeV)
in {\bf I} is due to including a consideration on the
correlation between $M_\sigma$ and $r_c$ in the present work,
and also, regrettably, due to mis-reading of the errors of some data
points of (ii) in {\bf I}.} 
with $\chi^2$=23.6 for 30 degrees of freedom
(34 data points with 4 parameters
\footnote{In (i), one point at $M_{\pi\pi}$=910 MeV 
seems to have a too small error and
to disturb the continuity to adjacent data points.
In fact, this point turned out to occupy large fraction of total $\chi^2$. 
The values in the text are obtained without this point,
while $M_\sigma$=600 MeV, $g_\sigma^r$=3750 MeV,
$r_c$=2.75 GeV$^{-1}$ (0.54 fm),
and $\sqrt{s_0}$=475 MeV with $\chi^2$=35.9 for 31 degrees of freedom,
in the case of including this point.}).
Note that the best-fitted core radius is nearly the same
as the ``structural size" (charge radius) of pion$\sim$0.7 fm.

In order to justify the preciseness of these values and
to estimate their errors it is apparently necessary to take into account 
of the correlation between $M_\sigma$, $g_\sigma$ and $r_c$.
Several fits are performed for various fixed values of $r_c$ 
between 0 and 4.0 GeV$^{-1}$. In Fig.2(b) $\chi^2$, $M_\sigma$ and
$g_\sigma$ are plotted as functions of $r_c$, where
$M_\sigma$ and $g_\sigma$ decrease as $r_c$ becomes larger.
The $\chi^2$ shows deep parabolic shape, and gives its minimum at 
3.03 GeV$^{-1}$, where we obtain the above mentioned ``best-fit'' values.
Note that the fit in the case of $r_c$=0, which corresponds to the results of
the conventional analyses without $\delta_{BG}$ thus far made,
gives $\chi^2$ = 163.4, about 140 worse than the best fit.
In the first row of Table 1 the errors of relevant parameters are quoted 
which correspond to three sigma deviation in 
the $\chi^2$ behavior in Fig.2(b).
In Fig.2(c) the fits with $r_c$=2.6 and 3.35 GeV$^{-1}$
corresponding to the 3-sigma deviations are shown. 
The small bump around 750 MeV of the standard 
$\delta_0^0$ is not reproduced with small 
$r_c$ (correspondingly with large $g_\sigma$ - width).
To estimate the effect due to the ambiguities of 
experimental $\delta^0_0$ data mentioned above,
data (v) and (vi) are analyzed with fixed $r_c$=3.03 GeV$^{-1}$.
The results are also shown in Fig.2(c) and the obtained 
values of parameters are given in the second and the third rows 
of Table 1.

From all of these studies, we may conclude that 
the $M_\sigma$ is in the range of {\bf 535-650 MeV}.

\section{Supplementary discussions}
\label{sec:dis}
We would add some comments on the results of our analysis:\\
\\
i) First a possible influence of the tail of $f_0(1370)$ 
even below $m_{\pi\pi}$=980 MeV should be mentioned. 
\footnote{
$f_0(1370)$ is known to couple strongly with $\pi\pi$ channel
and thus to give $180^\circ$ contribution to $\delta^0_0$.} 
Would $f_0(1370)$ have a large width, the value  
$r_c$=3.03 GeV$^{-1}$ in Table 1 should be regarded as including
the $f_0(1370)$ contribution.
\footnote{
In {\bf I} we obtained the relevant values 
$M_{f_0}$=1310 MeV, $\Gamma^{(p)}_{tot}$=420 MeV and
$r_c$=3.7 GeV$^{-1}$(0.73 fm).
However, as Svec96\cite{rf:Sv} and Kami\'nski96\cite{rf:Ka96}
 suggest, the analyses to obtain
$\delta_0^0$ thus far made do not usually take $a_1$ exchange
into account, which might affect $\delta^0_0$ in the mass region 
above $m_{\pi\pi}\sim$1 GeV. Also the width of $f_0(1370)$
is not actually known (see PDG96\cite{rf:PDG96}).
Exact estimation seems to be difficult at this stage.}
However, we may expect that it does affect very little 
the values of $M_\sigma$ and $g_\sigma$, since
these are determined mainly by the fine structure of the standard 
$\delta_0^0$ around 750 MeV.\\
\\
ii) In our treatment the S-matrix is parametrized directly 
in terms of physically meaningful quantities,
the masses and $\pi\pi$-coupling constants of resonances.
So it is not necessary to argue about pole 
positions on analytically continued complex Riemann sheets.
However, for convenience to compare with other works,
we give these values for the $\sigma$ resonance in Table 2.
Our S-matrix has the form of product of respective resonance-S-matrices.
Each BW resonance (in our form of Eq.(\ref{eq:bwb})) produces three poles,
a virtual state pole, a BW pole and its complex conjugate pole.
We are able to derive the approximate relations.
$Re \sqrt{s_{\rm pole}}$
$\approx M_R$ and 
$Im \sqrt{s_{\rm pole}}$
$\approx 1/2\Gamma^{(p)}$.
This situation corresponds to the case of one-channel ($\pi\pi$).
In the two-channel case ($\pi\pi$ and $K\bar{K}$), 
one BW resonance induces four poles (except for virtual state pole), 
$s_p^{III}$, its complex conjugate $s_p^{III*}$, and the two 
in sheet {\em II} or {\em IV}.
In {\bf I}, $g_{\sigma K\overline{K}}\approx 0$
is obtained. This case leads to the four poles with the same $s$-values
$s_p$ and $s_p^*$ on sheets {\em II} and {\em III}.

\begin{table}[t]
\caption{
Pole positions on sheet {\em II} in present one-channel analysis.
The errors corresponds to the 3-sigma deviation from the 
best fit to standard $\delta_0^0$.
}
\begin{center}
\begin{tabular}{|l|l|l|}
\hline
$s_{\rm pole}/{\rm  GeV}^2$ & $\sqrt s_{\rm pole}/$GeV
                                          & $p_{\rm pole}/$GeV\\
\hline 
 (0.324$\pm$0.020) &(0.602$\pm$0.026) &(0.271$\pm$0.015) \\ 
-i(0.236$\pm$0.044) & -i(0.196$\pm$0.027) &-i(0.109$\pm$0.014) \\
\hline
\end{tabular}
\end{center}
\end{table}

Here the decay width $\Gamma^{(d)}$ should be discriminated
from the peak width $\Gamma^{(p)}$. The former is defined by the formula
\begin{eqnarray*}
\Gamma^{(d)}&=& N^{-1}\int ds\Gamma (s)/[(s-M_\sigma^2 )^2+s\Gamma (s)^2];\\
N&=&ds\int 1/[(s-M_\sigma^2 )^2+s\Gamma (s)^2].
\end{eqnarray*}
\\
\noindent iii) 
In our analysis we have introduced 
a parameter $s_0$ in Eq.(\ref{eq:width}).
The obtained value of it is $\sqrt s_0=365$ MeV, 
which corresponds to the scattering
length $a_0^0=0.23$ in $m_{\pi}^{-1}$ unit.
We have introduced no ad hoc
prescriptions for Adler-zero condition. However,
we have found that our resultant amplitude has a zero
at $s=1.0m_\pi^2$.\\
\\
\noindent iv)
In the $I=2$ $\pi\pi$ system, there is
no known/expected resonance, and accordingly it is expected that
the repulsive core will show up itself directly.
In Fig.2 the experimental
data\cite{rf:BG} of the $I=2$ $\pi\pi$-scattering S-wave phase shift
$\delta_0^2$ is shown from threshold to $m_{\pi\pi}\approx 1400$ MeV,
which is apparently negative, and is fitted well also by the
hard core formula
$\delta_0^2=-r_c^{(2)}|{\bf p}_1|$ with the core radius of
$r_c^{(2)}=0.87\ {\rm  GeV}^{-1}(0.17\ {\rm fm})$.
The core radius is smaller by a factor of 3 than that in the $I=0$
system, which suggests the importance 
of quark-pair-annihilation force 
in the $\pi\pi$ system (see Ref.\citen{rf:YITP} for possible explanation 
in the framework of the linear $\sigma$ model).
\begin{figure}[t]
 \epsfysize=7.0 cm
 \centerline{\epsffile{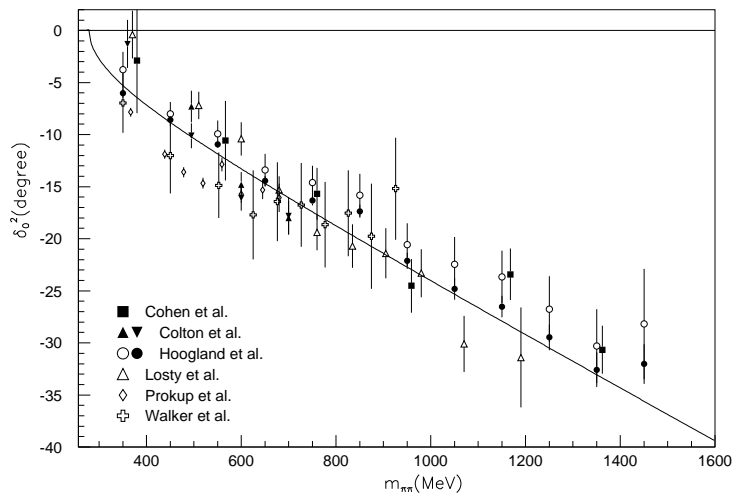}}
\caption{I=2 $\pi\pi$ scattering phase shift. Fitting by
hard core formula is also shown.}
\label{fig:i2}
\end{figure}

\section{Concluding remarks}
\label{sec:rem}
In this work and in {\bf I} we have shown the strong possibility
of existence of $\sigma$-particle with light mass ($\approx$580 MeV)
and comparatively narrrow width ($\approx$350 MeV).
The value of $M_\sigma$ is close to $M_\sigma\approx 2m_q$ ($m_q$
being constituent quark mass) expected in the NJL type of models.
Recently one of the present authors has argued\cite{rf:MY}
that our obtained values\footnote{
Recently Harada {\em et\ al.}\cite{rf:Ha} have made a similar 
analysis of $\pi\pi$-scattering data as {\bf I} leading to
the $\sigma$-existence with similar values of mass and width.
However, they start from the viewpoint of Non-linear $\sigma$
model and do not recognize it as a chiral partner of
$\pi$ meson.
}
of $m_\sigma$ and $\Gamma_\sigma$ are consistent with the
relation predicted in the linear $\sigma$-model (L$\sigma$M).
This fact seems to us to show that our ``observed" $\sigma$-particle
is really a chiral partner of $\pi$-meson as a Nambu-Goldstone
boson.

Here we also note an interesting argument\cite{rf:MY,rf:YITP} 
that the origin of 
repulsive core corresponds (,at least in the low 
energy region where the structure of composite hadrons is negligible,) 
to the ``compensating'' $\lambda\phi^4$ 
contact interaction, which is required from
the current algebra and the PCAC.\cite{rf:weinb,rf:yanagi}
\footnote{
The ``effective'' $\sigma\pi\pi$ coupling, which includes both 
effects of an intermediate $\sigma$-production and of the 
repulsive $\lambda\phi^4$ interaction, becomes of a derivative type,
while in the conventional Breit-Wigner formula, a non-derivative
coupling of $\sigma$ resonance is supposed.
This seems to be a reason
why $\sigma$-meson existence has been overlooked for many years.}

Historically the existence of $\sigma$-particle, although it was
anticipated from various viewpoints,\cite{rf:Sca,rf:HK,rf:KT,rf:MLS}
had been rejected for many years. One of the main reasons was that
(i) the $\sigma$ has been missing in the $\pi\pi$ phase shift 
analyses,\cite{rf:MP} and the other was due to (ii) the 
negative results of applications\cite{rf:GL} of L$\sigma$M 
to the low energy $\pi\pi$ scattering and to the $K_{l4}$ decay 
form factors.

Concerning to (i), the present work may give a possible and clear 
solution. Recently there are several other phase shift 
analyses\cite{rf:Ka,rf:Tor,rf:Ha} resulting in $\sigma$-particle existence. 
Also the problem (ii) should be reexamined\cite{rf:MY} under the light 
of these recent
progress of phenomenological search for $\sigma$-particle.
\footnote{
Ref.\citen{rf:shaba2} argues that the L$\sigma$M describes 
successfully both of the mentioned processes.
}
Finally we should like to note that recently a rather 
strong evidence\cite{rf:Taku}
for direct $\sigma$ production has been obtained in the central
{\em pp}-collision. As a matter of fact, 
it was the motivation of our investigation.

\end{document}